\documentclass[paper=a4, fontsize=11pt]{scrartcl}
\usepackage[T1]{fontenc}
\usepackage[utf8]{inputenc}
\usepackage{fourier}

\usepackage[english]{babel}															
\usepackage[protrusion=true,expansion=true]{microtype}	
\usepackage{amsmath,amsfonts,amsthm,dsfont} 
\usepackage[pdftex]{graphicx}	
\usepackage{url}

\usepackage{sectsty}
\allsectionsfont{\centering \normalfont\scshape}

\usepackage{fancyhdr}
\usepackage{caption}
\usepackage{hyperref}
\newcommand{\markov}{\mathcal{M}}
\newcommand{\prob}{\mathds{P}}
\newcommand{\probm}{\prob_{\markov}}
\newcommand{\indep}{\perp\!\!\!\perp}
\renewcommand{\natural}{\mathds{N}}
\newcommand{\LG}{\mathcal{L}_G}
\newcommand{\LnG}{\mathcal{L}_G^n}
\newcommand{\A}{\mathcal{A}}
\newcommand{\V}{\mathcal{V}}
\newcommand{\C}{\mathcal{C}}
\newcommand{\OO}{\mathcal{O}}
\newcommand{\one}{\mathds{1}}
\newcommand{\derive}{\rightarrow^*}

\newtheorem{definition}{Definition}

\numberwithin{equation}{section}		
\numberwithin{figure}{section}			
\numberwithin{table}{section}			

\newcommand{\horrule}[1]{\rule{\linewidth}{#1}} 	

\title{
		\usefont{OT1}{bch}{b}{n}
		\normalfont \normalsize \textsc{Sony Computer Science Laboratory} \\ [25pt]
		\horrule{0.5pt} \\[0.4cm]
		\huge Sampling Markov Models under Constraints: Complexity Results for Binary Equalities and Grammar Membership \\
		\horrule{2pt} \\[0.5cm]
}
\author{
		\normalfont \normalsize
        Stéphane Rivaud \\
        \small
        email: \href{mailto:stephane.rivaud@sony.com}{stephane.rivaud@sony.com} \\
		\normalsize
        François Pachet \\
        \small
        email: \href{mailto:pachetcsl@gmail.com}{pachetcsl@gmail.com} 	\\
		\tiny
        \date{\today}
}

\begin{document}
\maketitle

\begin{abstract}
We aim at enforcing hard constraints to impose a global structure on sequences generated from Markov models.
In this report, we study the complexity of sampling Markov sequences under two classes of constraints: Binary Equalities and Grammar Membership Constraints.
First, we give a sketch of proof of \#P-completeness for binary equalities and identify three sub-cases where sampling is polynomial.
We then give a proof of \#P-completeness for grammar membership, and identify two cases where sampling is tractable.
The first polynomial sub-case where sampling is tractable is when the grammar is proven to be unambiguous.
Our main contribution is to identify a new, broader class of grammars for which sampling is tractable. 
We provide algorithm along with time and space complexity for all the polynomial cases we have identified.
\end{abstract}

%
%

\section{Introduction}

Markov models are widely used in probabilistic sequential data modeling.
Their associated graphical model is a linear chain which is a simple topology, allowing inference tasks like computing marginals to be achieved in polynomial time. They are easy to implement and can be  trained to fit perfectly a dataset in one pass.
This simplicity makes them suitable for many tasks such as automatic content generation.

However, the Markov assumption underlying these models is quite restrictive.
For a $d$-th order Markov chain $s=s_1,\cdots,s_{n}$, the assumption states that the probability of a value at position $i$ in a sequence only depends on the $d$ preceding values:
\begin{align}
\prob(s_i | s_1, s_2, \cdots, s_{i-1})= \prob(s_i | s_{i-d}, \cdots, s_{i-1})
\end{align}

A classical way to increase the expressivity of Markov models is to increase the order, but this comes at an exponential cost in time and space complexity along with over-fitting issues when the training set is small. As a result, sequences generated from Markov models usually lack global structure and are not able to capture long range dependencies.

In order to address this issue, we investigate the use of hard constraints to enforce such global properties at the sampling phase on the generated sequence.
Our goal is to express these properties as a Constraint Satisfaction Problem (CSP), and to sample solutions of the CSP with the probability given by the Markov model.
Several results have been obtained toward this aim \cite{pachet2015generating, papadopoulos2015exact, papadopoulos2016generating, papadopoulos:14a}. 

In this paper, we review complexity results previously obtained on the sampling of Markov sequences constrained with binary equalities \cite{rivaud:16a}. We also extend previous work on the Regular constraint \cite{papadopoulos2015exact,papadopoulos2016generating,roy2013enforcing} by investigating a more expressive constraint, the grammar membership constraint \cite{sellmann2006theory}, and give complexity results for sampling under such constraints.

\section{Sampling Markov Sequences under Binary Equality Constraints}

\paragraph{Homogeneous Markov Model:} Let $X_1, X_2, \hdots$ be a sequence of random variables taking values in a set $\A$. A homogeneous Markov model $\markov$ consists of a transition matrix $T_{\markov}$ and a vector $P_0$ such that:
\begin{itemize}
	\item $T_{\markov}(i, j) = \prob( X_{k+1} = i | X_k = j )$
    \item $P_0(i) = \prob(X_0 = i)$
\end{itemize}
The Markov assumption underlying a Markov model states that for all integers $i, j$ and $k$ such that $0 \leqslant i < k < j$, we have  $X_i \indep X_j | X_k$, i.e. given $X_k$, $X_i$ and $X_j$ are conditionally independent. \\

\paragraph{Binary Equality Constraint:} A binary equality constraint between two variables $X_i$ and $X_j$ taking values in $\A$ is satisfied whenever $X_i = X_j$.
These constraints are easily filtered and solutions can be uniformly sampled efficiently.
We can extend the binary equalities to any binary relation where the value of $X_i$ determines the value of $X_j$ and vice-versa, e.g. $X_i = \sigma(X_j)$ where $\sigma$ is a permutation of the set $\A$.
These are particularly useful when one wants to obtain specific patterns in the sequences sampled from Markov models.
However, even if the problem is easily solved when the random variables are independent, it becomes \#P-complete when imposing a Markov distribution on the sequence.
The complete proof as well as some examples of chord sequence generation can be found in \cite{rivaud:16a}. Here we give a sketch of the proof for the problem of sampling Markov sequences subject to binary equality constraints.\\

In this paper, we will assume that a sampling problem associated with a CSP is equivalent to its (approximate) counting version.\\

Counting the number of solutions in a binary CSP (and more generally pairwise Markov network) is \#P-hard.
Our proof in \cite{rivaud:16a} consists in building a reduction from the problem of counting the number of solutions of a binary CSP to the problem of sampling Markov sequences constrained by binary equality constraints.
The idea is to "unwrap" the factor graph of the binary CSP into a linear chain constrained with binary equalities, and to normalize the factors into a Markov transition matrix.
Thus, any binary CSP can be seen as the contraction of a Markov model with binary equalities. \\

Let $G$ be a factor graph associated to a binary CSP. The algorithm works in 3 steps:
\begin{enumerate}
\item Add dummy edges to make the graph $G$ Eulerian.
\item Find an Eulerian path for $G$ and build a Markov model $\markov$ out of this path.
\item Impose equality constraints between variables corresponding to the same vertex.
\end{enumerate}

\begin{figure}[h]
	\begin{center}
		\includegraphics[scale=0.85]{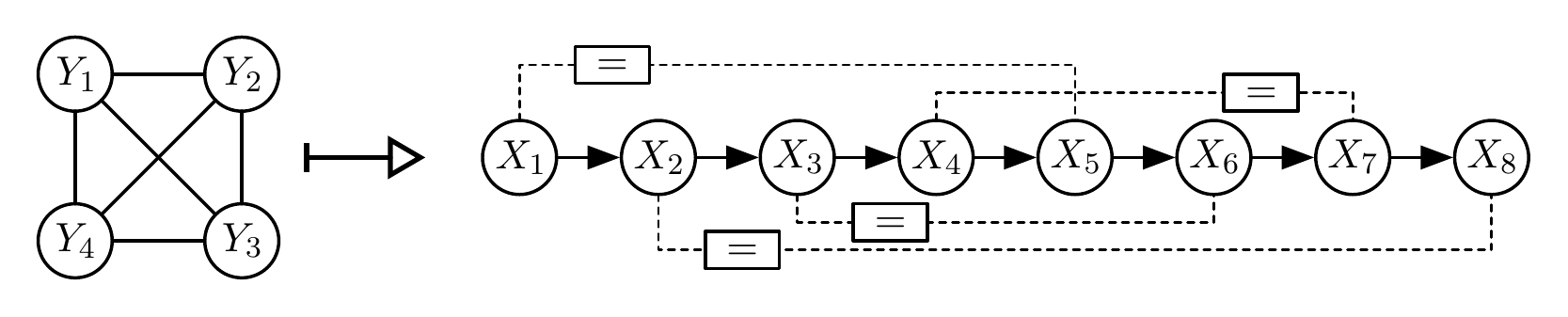}
        \caption{On the left is a factor graph representing a binary CSP.\\
        On the right is the equivalent Markov model with a set of equalities.}
	\end{center}
\end{figure}

We can show that if one is able to perform perfect sampling with respect to a Markov model subject to binary equality constraints, he can also count the number of solutions of a binary CSP, which establishes the \#P-completeness of the problem.

However, we can identify specific cases where inference is tractable by looking at the topology of the quotient graph. Three cases can be useful in practice:
\begin{itemize}
\item \textbf{Non-crossing equalities} when the variables inside the equalities do not overlap. The time complexity is in $\OO(|\A|^2 n)$ with $n$ being the length of the sequence.
\item \textbf{Repeated sections} when, given a list of binary equalities among variables, the topological order of the first variable for each equality is the same as for the second variable for each equality. The time complexity is in $\OO(|\A|^3 n)$.
\item \textbf{Palindromic sections} when the set of equalities can be written $\{X_{i_1} = X_{j_1}, \hdots, X_{i_K} = X_{j_K}\}$ with $i_1 < \hdots < i_K < j_K < \hdots < j_1$. The time complexity is in $\OO(|\A|^2 n)$.
\end{itemize}

\begin{figure}[h!]
	\begin{center}
		\includegraphics[scale=0.8]{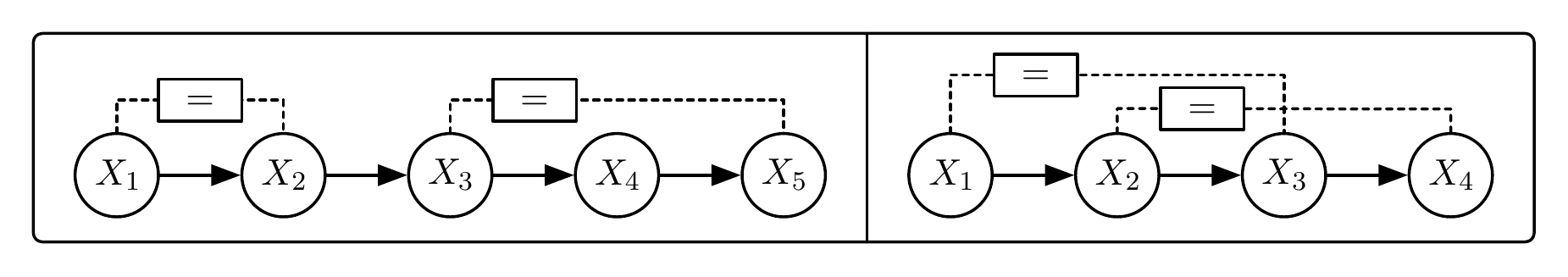}
        \caption{On the left, a non-crossing equalities configuration. On the right, a repeated section configuration.}
	\end{center}
\end{figure}

\begin{figure}[h!]
	\begin{center}
		\includegraphics[scale=0.35]{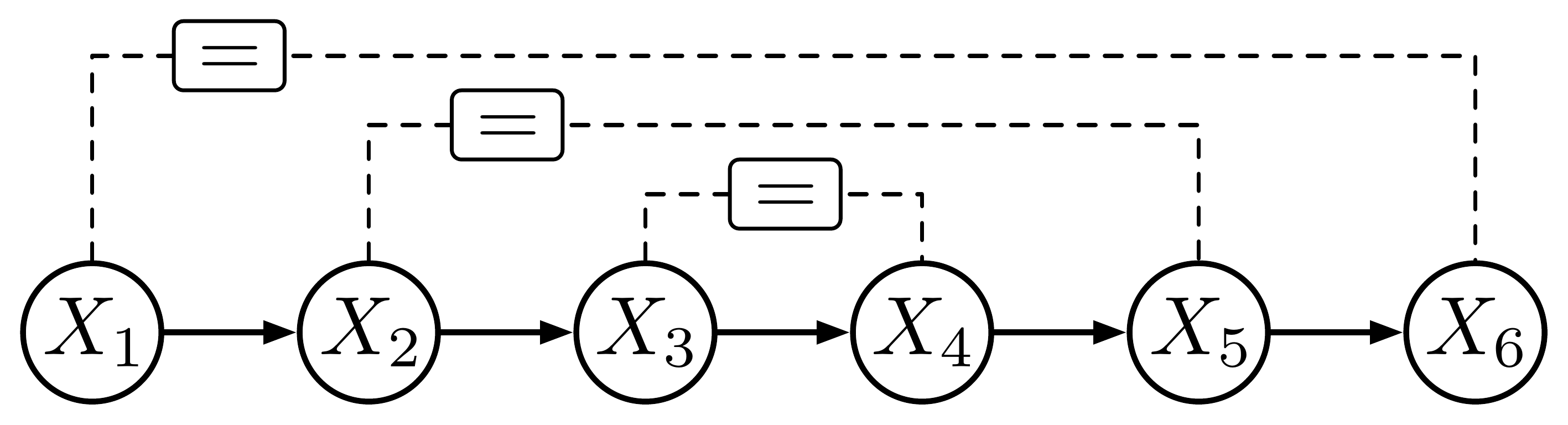}
        \caption{Graph representing the palindromic case of binary equalities with Markov models.}
	\end{center}
\end{figure}

%
%

\section{Sampling Syntactically Correct Markov Sequences}

Algebraic languages, or equivalently Context-Free Grammars (CFGs) form the highest level in the Chomsky hierarchy for which the word problem is polynomial, i.e. the problem of deciding if a word $w$ is generated by a given grammar $G$.
The use of algebraic languages in Constraint Programming was studied in \cite{sellmann2006theory,quimper2006global}, with discussions on the efficiency of implementation in \cite{kadioglu2008efficient,katsirelos2009restricted}.

To deal with probabilities, a classical extension of CFGs are Probabilistic CFGs (PCFGs), which define a probability distribution on the set of words of length $n$. Another extension, Weighted Context-Free Grammars, has been proposed in \cite{katsirelos2011weighted}. In this approach, weights are assigned to each parsing tree without the restriction that it must define a probability distribution, but it has been proven to be as expressive as PCFGs \cite{smith2007weighted}.
There have been numerous applications of PCFGs in many domains \cite{geman2002probabilistic} including Natural Language Processing \cite{sundararajanprobabilistic}, secondary structure discovery in proteins \cite{interesting2001stochastic} and automatic music generation \cite{groves2016automatic,mcleod2017meter,nakamura2016tree,abdallah2014exploring}.
However, training of these models is difficult due to ambiguity issues \cite{sima2002computational} and approximate iterative methods are often used in practice.
This situation makes them unsuitable for on-line learning and incremental sampling where we have to sample one variable at each time-step, which is not the case when dealing with Markov models.
On the other hand, \emph{Regular} constraints can easily be expressed in the framework of grammar membership constraints.
Having one global framework to perform constrained sampling would allow a user to specify a wide variety of constraints without changing the implementation.
This is typically useful in a work-flow when one would like to implement an additional control tool on a user interface without  modifying the underlying engine \cite{papadopoulos2016assisted}.
We therefore investigate the complexity of imposing a grammar membership constraint on a Markov model. \\

More formally, a Context-Free Grammar (CFG) is a quadruplet $G = (\V, \A, R, S_0)$ where
\begin{itemize}
	\item $\V$ is a set of variables called non-terminals.
    \item $\A$ is a set of variables called terminals, disjoint from $V$.
    \item $R \subset \V \times (\V \cup \A)^*$ is a set of production rules.
    \item $S_0 \in \V$ is the initial symbol.
\end{itemize}
It is a rewriting system used to generate words by starting with the initial symbol $S_0$, and recursively applying rules from $R$ to the letter of the word until it belongs to $\A^*$.
Each word generated with this process inherently possesses a parse-tree representing the sequence of rules used to generate it.
In the sequel, we give a proof that the general case is \#P-complete, and we investigate two sub-cases where inference is tractable and give polynomial algorithms for these cases.

\subsection{General case is \#P-complete}
Performing inference tasks like computing marginals or the normalization constant with respect to a model $\markov$ subject to a constraint $\C$ is harder than counting the number of assignments satisfying the constraint $\C$. We show that counting the number of strings of an ambiguous grammar is \#P-complete, even when the given grammar is Regular. An ambiguous Regular grammar can be represented with a non-deterministic finite automaton (NFA). We show that by reducing the \#2SAT enumeration problem to the one of counting the strings of a given length $n$ accepted by a NFA.\\

Let $\varphi$ be a 2SAT Boolean formula over variables $x_1, \hdots, x_n$ in conjunctive normal form,
i.e. $\varphi = \varphi_1 \wedge \hdots \wedge \varphi_K$ with $\varphi_k = l^k_{i_1} \vee l^k_{i_2}$ for all $k \in [1, K]$ and $l^k_i$ is a literal equal to a variable or its negation.
We denote by SAT($\varphi$) the set of assignment that satisfy $\varphi$ and UNSAT($\varphi$) the ones who falsify $\varphi$.
As there is $2^n$ possible assignment, we have that \#SAT($\varphi$) = $2^n$ - \#UNSAT($\varphi$).
We will build an NFA exactly recognizing sequences in UNSAT($\varphi$), thus counting these sequences is equivalent to counting \#UNSAT($\varphi$) and thus \#SAT($\varphi$).\\

\paragraph{A falsifying NFA: }
Let $s_0$ denote the initial state.
For each clause $\varphi_k = l^k_{i_1} \vee l^k_{i_2}$, we add $n$ state $s^k_1, \hdots, s^k_n$.
We then add a transition $(s_0, s^k_1)$, and the transitions $(s^k_{i-1}, s^k_i)$ for all $i \in [2, N]$ with the following label:
\begin{itemize}
	\item 	if $i_1 = 1$ then if $l^k_{i_1}$ is a positive literal, then $(s_0, s^k_1)$ is labeled with 0, otherwise it is labeled with 1.
    		if $i_1 \neq 1$ then $(s_0, s^k_1)$ is labeled with 0 and 1.
	\item 	if a literal $l^k_i$ is positive then the transition $(s^k_{i-1}, s^k_i)$ is labeled with 0, otherwise it is labeled with 1.
	\item 	the remaining transitions are labeled with 0 and 1.
\end{itemize}
We then set $s^k_n$ for all $k \in [1, K]$ as acceptance states.
If a word $x_1 \hdots x_n$ reaches a state $s^k_n$, then it falsifies the clause $\varphi_k$ and thus $\varphi$.
Therefore, the number of strings of length $n$ accepted by the constructed automaton is the number of assignments in UNSAT($\varphi$).

\begin{figure}
	\begin{center}
		\includegraphics[scale=0.40]{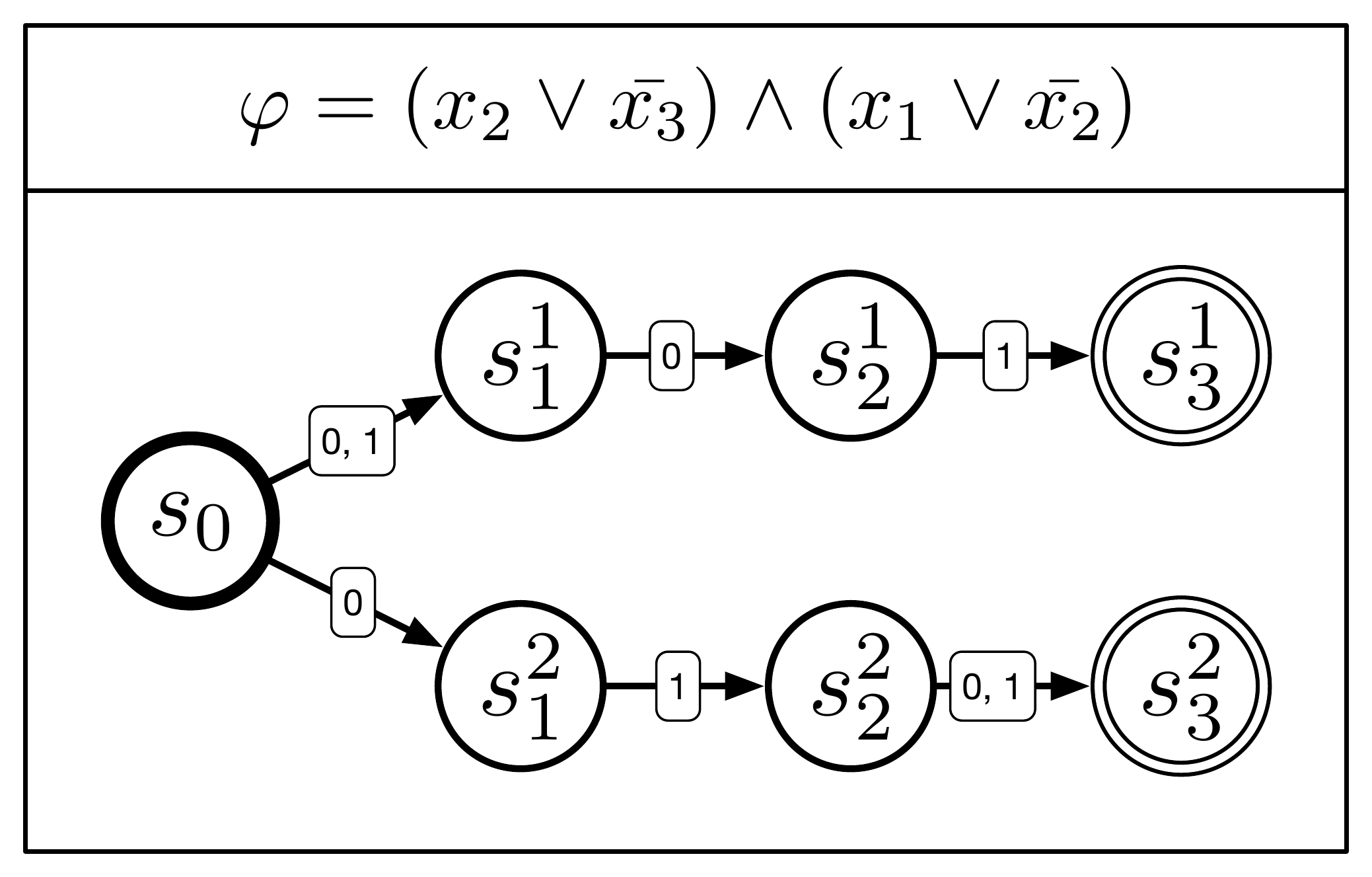}
		\caption{On the top is a 2-SAT formula.\\
        On the bottom, its associated falsifying NFA.}
	\end{center}
\end{figure}

\paragraph{Complexity :}
The automaton built with our procedure comprises 1 initial state and $n$ additional states per clause.
There are $K$ clauses, thus the complexity of the reduction is $\OO(Kn)$, which is linear in the size of the formula.
Since counting the number of satisfying assignments for a 2SAT formula is \#P-complete \cite{valiant1979complexity}, then counting the number of strings of length $n$ accepted by an NFA is \#P-complete, and so is the problem of counting the number of strings of length $n$ generated by an ambiguous grammar.\\

It is worth noting that counting the number of strings accepted by a deterministic finite automaton can easily be done in polynomial time, with message passing algorithms for example. Therefore, the non-deterministic aspect of the automaton, which translates into ambiguity in the corresponding Regular grammar makes the counting problem \#P-complete. In the sequel, we investigate restricted classes of context-free grammars where the counting problem can be answered in polynomial time. We then give an algorithm to perform common inference tasks with respect to a Markov model subject to a grammar membership constraint.

\subsection{The unambiguous case}

One of the most common inference task is to compute the normalization constant of a distribution. For example, given $n \in \natural$, a grammar $G$ and its associated language $\LG$, it is useful to compute $\prob_{\markov}(\LnG)$ where $\LnG$ is the set of words of length $n$ that belong to $\LG$, i.e. $\LnG = \{ w \in \A^n \;|\; w \in \LG , \; |w| = n\}$. In order to do so, we need to decompose the probability in a computationally efficient way. We therefore introduce some notations and additional variables. \\

\paragraph{Hidden variables:}
Given a non-terminal symbol $A \in \V$ and a word $w \in \A^*$, we note $A \derive w$ if we can derive $w$ from the non-terminal $A$. We note $\LG(A)$ the set of words that can be derived starting from $A$, i.e. $\LG(A) = \{ w \in \A^* \; | \; A \derive w \}$. We also denote by $\LnG(A)$ the set of word  of length $n$ in $\LG(A)$. We introduce the random variables $S_i^j$ for $j$ from 1 to $n$, $i$ from 1 to $n-j+1$ and $S \in \V$ taken from the filtering algorithm presented in \cite{sellmann2006theory}. These binary variables follow  a conditional distribution  $\probm(V_i^j = 1 \; | \; X_i, \hdots, X_{i+j-1})$ which is equal to 1 if $V \derive X_i \hdots X_{i+j-1}$ and 0 otherwise. We therefore have:

$$
\probm(V_i^j = 1) = \sum_{X_i, \hdots, X_{i+j-1}} \probm(X_i, \hdots, X_{i+j-1}) \one_{\LG^j(V)}(X_i, \hdots, X_{i+j-1}) 
$$

It is important to note that for $A, B \in \V$ such that $A \neq B$, then $A_i^j$ and $B_i^j$ are different random variables, i.e. $\sum_{V \in \V} \probm(V_i^j) \neq 1$.
Each of these random variables partitions the space $\A^n$ into two sets, namely $\{V_i^j = 1\}$ and $\{V_i^j = 0\}$.
In the sequel, we will write $\probm(V_i^j)$ in lieu of $\probm(V_i^j = 1)$ to denote the probability of the set.

\paragraph{Dynamic Programming:}
With these notations, we can decompose the probability with the following recurrence formula:
\begin{equation}
\probm(V_i^j) = \sum_{k = 1}^{j-1} \; \sum_{ (V \rightarrow A B) \in \V} 	\probm(V_i^j \; | \; A_i^k, B_{i+k}^{j-k}) 	\probm(A_i^k, B_{i+k}^{j-k})
\label{rulesDecomposition}
\end{equation}
where $\probm(V_i^j \; | \; A_i^k, B_{i+k}^{j-k}) = 1$ if and only if $V \rightarrow A B$ is a valid production rule.
This decomposition basically computes the weighted sum of the trees rooted in $V_i^j$ where each tree is associated to the Markov probability of its corresponding word.
As the grammar is unambiguous, there is a one-to-one mapping between the set of parse trees and the set of words in $\LG$.

\paragraph{Markov dependencies:}
The last quantity we need to compute is $\probm(A_i^k, B_{i+k}^{j-k})$.
It represents the probability that $A \rightarrow^* X_i \hdots X_{i+k-1}$ and $B \rightarrow^* X_{i+k} \hdots X_{i+j-1}$.
If the random variables $(X_1, \hdots, X_n)$ were independent, we would simply have $\probm(A_i^k, B_{i+k}^{j-k}) = \probm(A_i^k)\probm(B_{i+k}^{j-k})$.
In the Markov case, we use the Markov assumption to keep the decomposition tractable.
Namely, we have that given $X_{i+k-1}$ or $X_{i+k}$, the variables $A_i^k$ and $B_{i+k}^{j-k}$ are conditionally independent:
\begin{equation}
\probm(A_i^k, B_{i+k}^{j-k}) = \sum_{X_{i+k}} 		\probm(B_{i+k}^{j-k} | X_{i+k})
                            	\sum_{X_{i+k-1}} 	\probm(X_{i+k} | X_{i+k-1})
													\probm(X_{i+k-1} | A_i^k)
                                					\probm(A_i^k)
\label{markovTransition}
\end{equation}

\paragraph{Complexity:}
Different trade-offs can be obtained between time and space complexity.
According to our decomposition, we need to store $\probm(V_i^j | X_i)$ and $\probm(V_i^j | X_{i+j-1})$ which costs $\OO(n^2 |G|^2)$.
We can also store the marginals $\probm(V_i^j)$ at the cost of $\OO(n^2 |G|)$ without increasing the asymptotic space complexity. The matrix $\probm(X_{i+k} | X_{i+k-1})$ is provided by the input as it is part of the Markov model, and the question to store the marginals $\prob(X_{k})$ is left to the trade-off between time or space efficiency.\\

The time complexity can be read through the different summations we need to perform in equation \eqref{rulesDecomposition} and \eqref{markovTransition}.
Equation \eqref{markovTransition} performs summation over $X_{i+k-1}$ and $X_{i+k}$. We usually perform summation over $X_{i+k-1}$, store the result for each values of $X_{i+k}$, and then perform summation over $X_{i+k}$.
This requires $\OO(|G|)$ in space but yields a double summation in time $\OO(|G|)$.
We apply \eqref{markovTransition} for all $i, j, k$ such that $1 \leq i < i + k < j \leq n$ which is $\OO(n^2)$.
Providing all entries to perform \eqref{rulesDecomposition} thus costs $\OO(n^2 |G|)$.

Given the necessary data, equation \eqref{rulesDecomposition} performs summation over all integers $k \in [i, j]$ and production rules $(V \rightarrow A B) \in G$ thus requiring $\OO(n |G|)$ for each variable $V_i^j$.
We apply equation \eqref{rulesDecomposition} for all $1 \leq i < j \leq n$, which represent $ \frac{n(n+1)}{2} = \OO(n^2)$ variables, resulting in $\OO(n^3 |G|^3)$ time complexity.
This dominates the complexity associated with \eqref{markovTransition} and gives a total time complexity of $\OO(n^3 |G|^3)$.

\begin{figure}
	\begin{center}
		\includegraphics[scale=0.37]{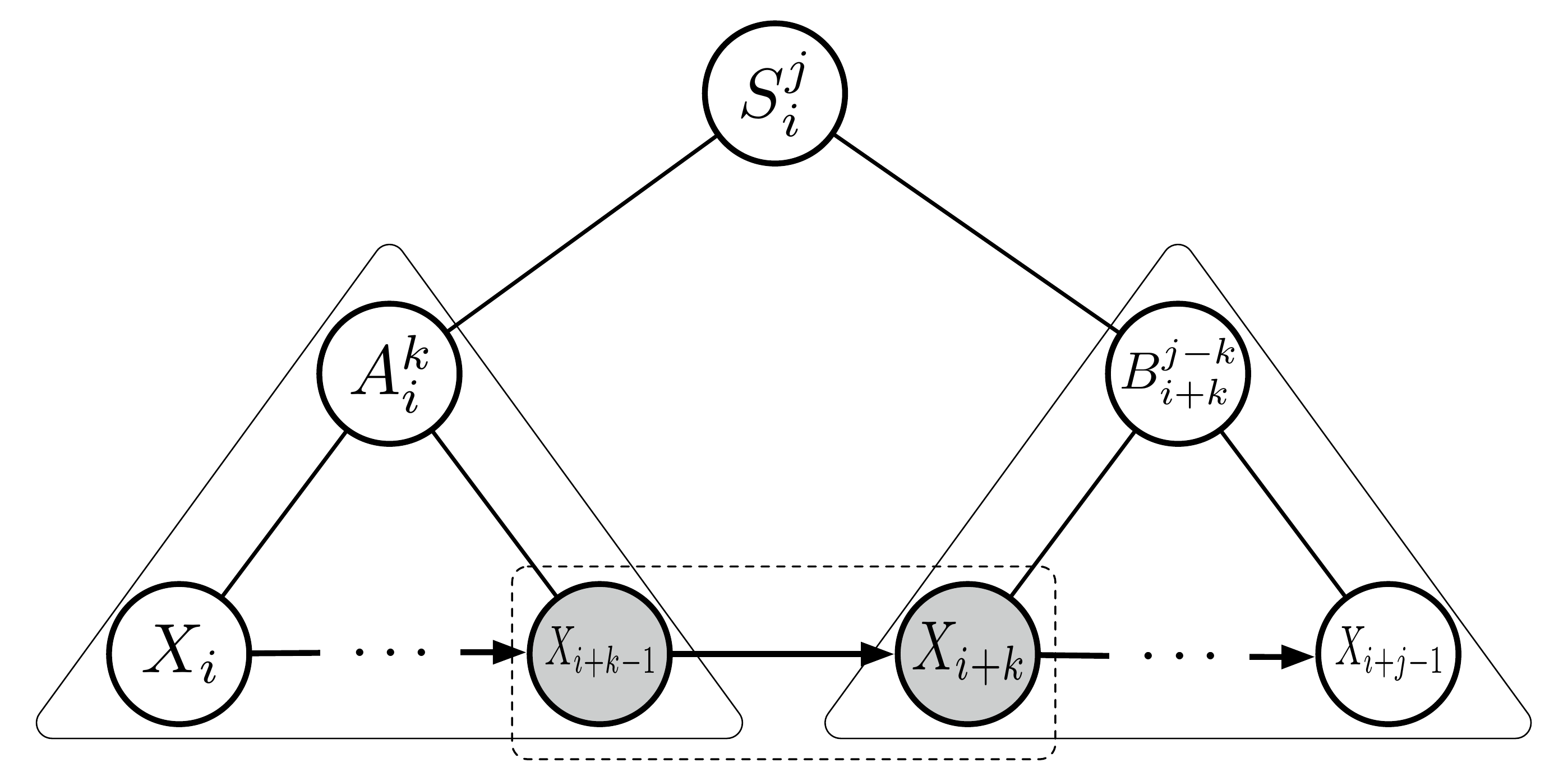}
        \caption{Graph representing the dependencies among variables. The arrows represent the Markov dependencies, and the lines represent the dependencies induced by the grammar constraint.}
	\end{center}
\end{figure}

\subsection{A weakly-ambiguous case}

As shown earlier, there is little hope to find a polynomial algorithm in the general ambiguous case, but we can relax the unambiguity hypothesis into a weakly-ambiguous hypothesis.

\begin{definition}
A context-free grammar $G$ is weakly ambiguous if the languages generated by its symbols in $\V$ are disjoint, i.e. $\LG(A) \cap \LG(B) = \emptyset$ for all $A , B \in \V$ such that $A \neq B$.
\end{definition}

This condition imposes that if a substring $x_i \hdots x_{i+j}$ appears in a word of length $n$, then there is only one symbol that could have generated it. Therefore, if there are two parsing trees associated with a word, then the two trees must "intersect" in some way. We formalize this idea of intersection in this section.

\paragraph{Ordering the trees:}
We denote by $\C_n$ the set of syntactic trees having $n$ leaves, $\C_n(S)$ the subset of trees in $\C_n$ that has a root labeled with $S \in \V$, $\C_n^k(S)$ the subset of trees in $\C_n(S)$ having their left sub-trees in $\C_k$ and $\C_n^k(S, A, B)$ the subset of trees in $\C_n^k(S)$ having their left sub-tree in $\C_k(A)$ and their right sub-tree in $\C_{n-k}(B)$. \\

The sets $\hat{\C_n}(S)$ for all $S \in \V$ (resp. $\C_n^k(S)$ for all $S \in \V$ and all $k \in \{1, \hdots, n-1\}$, $\C_n^k(S, A, B)$ for all $S, A, B \in \V$ and all $k \in \{1, \hdots, n-1\}$) partition the set of syntactic parse trees.
As there is a natural mapping from the set of parse trees into $\LG$, we can associate each set of trees to its corresponding sets of words.
However if the grammar is ambiguous they do not induce a partition on the set words generated by the grammar, because one word might be associated with multiple parse trees.\\

We build an order on syntactic trees with the following recursive procedure. First, we impose that for any $T_1 \in \C_p$ and $T_2 \in \C_q$, we have $T_1 < T_2 $ if $p < q$. Then, we choose an order on $\V \cup \A$, which can always be done because the set is assumed to be finite. Finally for $T_1, T_2 \in \C_n$, we have:
\begin{enumerate}
\item if $n = 1$, let's note $s_1$ and $s_2$ the label of the unique node in $T_1$ and $T_2$. We say that $T_1 < T_2$ if $s_1 < s_2$ for the order imposed on $\V \cup \A$.
\item if $n > 1$, we note $T_1 = (s_1, L_1, R_1)$ and $T_2 = (s_2, L_2, R_2)$ where $s_i, L_i, R_i$ are the label of the root in $T_i$, the left sub-tree, and the right sub-tree. We say that $T_1 < T_2$ if $(s_1, L_1, R_1) < (s_2, L_2, R_2)$ for the lexicographical order.
\end{enumerate}

\paragraph{Partitioning the words:}
Let us now consider a word $w \in \A^n$ and let us note $T(w)$ the set of parse trees associated with $w$.
As $T(w)$ is finite and ordered the application
$$\hat{T}: w \mapsto \min_{t \in T(w)} t$$
is well defined and induces a partition on $\LnG$ for any $n \in \natural^*$.
We denote by $\widehat{\C}_n(S)$ (resp. $\hat{\C}_n^k(S), \widehat{\C}_n^k(S,A,B)$) the set of words $w$ such that $\widehat{T}(w)$ belongs to $\C_n(S)$ (resp. $\C_n^k(S)$, $\C_n^k(S,A,B)$). With these notations, we can write the following partition:
$$
\LnG = \widehat{\C_n(S)} = \cup_{k=1}^n \cup_{A,B} \widehat{\C_n^k}(S,A,B)
$$

Considering $\C_n^k(S, A, B)$ as a set of words, it is composed of all the words that have a syntactic parse tree $T = (S, L, R)$ with $L$ in $\C_k(A)$ and $R$ in $\C_k(A)$. The set  $\widehat{\C_n^k}(S, A, B)$ is the set of words such that $\widehat{T}(w)$ belongs to $\C_n^k(S, A, B)$. We can thus write:

$$
\widehat{\C_n^k}(S, A, B) = \C_n^{k}(S, A, B) \; \backslash \; \C_n^{<k}(S, A, B)
$$
where $\C_n^{<k}(S, A, B))$ is the set of words $w$ in $\C_n^{k}(S, A, B)$ such that $\hat{T}(w)$ is not in $\C_n^{k}(S, A, B)$. We can decompose this set in the following way:
$$
\C_n^{<k}(S, A, B) = \C_n^k(S, A, B) \cap \left( \cup_{(k', C, D) < (k, A, B)} \widehat{\C_n^k}(S, C, D) \right)
$$
where $(k', C, D) < (k, A, B)$ is the lexicographical order.

\paragraph{Dynamic Programming:}
We want to compute $\probm(S_1^n)$. In order to decompose the probability in a tractable way, we use the previous notations:

$$
\probm(S_i^j) = \sum_{k = 1}^{j-1}  \sum_{(S \rightarrow A B) \in G} \probm(\widehat{\C_{i,j}^k}(S,A,B)) \\
$$

with 

$$
\prob_{\markov}(\widehat{\C_{i,j}^k}(S,A,B)) = \probm(S_{i}^{j}, A_{i}^{k}, B_{i+k}^{j-k}) - \probm(\C_{i,j}^{<k}(S,A,B))
$$

We have
$$
\probm(S_{i}^{j}, A_{i}^{k}, B_{i+k}^{j-k}) = 
\left\lbrace
	\begin{array}{ll}
		\probm(A_i^k, B_{i+k}^{j-k})	&\text{if} \quad S \rightarrow A B \quad \text{is a valid production rule} \\
		0 								&\text{otherwise} \\
	\end{array}
\right.
$$

and

$$
\probm(\C_{i,j}^{<k}(S,A,B)) = \sum_{(k', C, D) < (k, A, B)} \probm(\widehat{\C_{i,j}^{k'}}(S,C,D), \C_{i,j}^k(S,A,B))
$$

This last term can be computed with the following decomposition:
$$
\probm(\widehat{\C_{i,j}^{k'}}(S,C,D), \C_{i,j}^k(S,A,B)) = \sum_{E_{i+k'}^{k-k'}} \probm(S_i^j, C_i^{k'}, D_{i+k'}^{j-k'}, A_i^k, B_{i+k}^{j-k}, E_{i+k'}^{k-k'})
$$
This decomposition is correct since we assume that our grammar is weakly ambiguous.
Let us suppose there are two parse trees $T_1$ and $T_2$ associated to a word $x_1 \hdots x_n$ with $T_1 \in \C_n^{k'}(S,C,D)$ and $T_2 \in \C_n^k(S,A,B)$. If $k =k'$ then $A = C$ and $B = D$ due to the weak ambiguity. Let us suppose then that $k' < k$:
\begin{itemize}
\item The symbol generating $x_1 \hdots x_{k'}$ in $T_1$ and $T_2$ is $C$.
\item The symbol generating $x_k \hdots x_n$ in $T_1$ and $T_2$ is $E$.
\item Their is $E \in \V$ such that the $E$ is the symbol generating $x_{k'} \hdots x_k$ in both $T_1$ and $T_2$.
\end{itemize}
Therefore, to compute $\probm(\C_{i,j}^{<k}(S,A,B))$, we need to account for all trees in $\C_{i,j}^k(S,A,B)$ such that there is a tree in $\C_{i,j}^{k'}(S,C,D)$ fulfilling the above conditions for all $C,D \in \V$ and $k' < k$. When these conditions are satisfied, we have:
$$
\probm(S_i^j, C_i^{k'}, D_{i+k'}^{j-k'}, A_i^k, B_{i+k}^{j-k}, E_{i+k'}^{k-k'}) = \probm(C_i^{k'}, E_{i+k'}^{k-k'}, D_{i+k'}^{j-k'})
$$

\begin{figure}
	\begin{center}
		\includegraphics[scale=0.37]{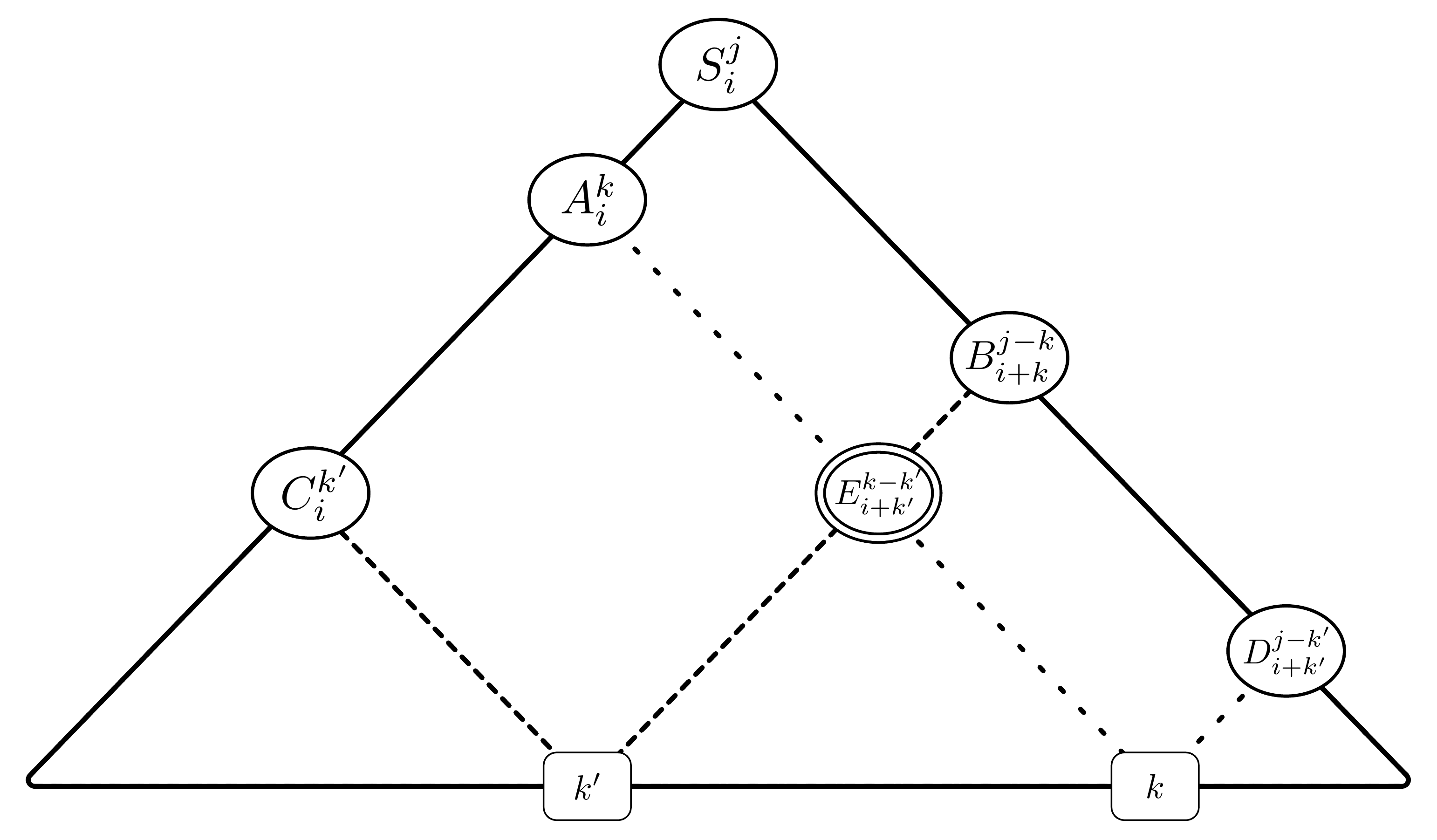}
        \caption{Graph representing the point how two parse trees representing the same word should intersect.}
	\end{center}
\end{figure}

\paragraph{Markov dependencies:}
The last quantities we need to compute are $\probm(A_i^k, B_{i+k}^{j-k})$ and $\probm(C_i^{k'}, E_{i+k'}^{k-k'}, B_{i+k}^{j-k})$, which takes into account the Markov dependencies among variables. The first term $\probm(A_i^k, B_{i+k}^{j-k})$ can be computed the same way it as in the unambiguous case.\\

With the same approach, we have: given $X_{i+k'-1}$ or $X_{i+k'}$, and $X_{i+k-1}$ or $X_{i+k}$, the variables $C_i^{k'}, E_{i+k'}^{k-k'}, B_{i+k}^{j-k}$ are conditionally independent. We can therefore write
\begin{align*}
\probm(C_i^{k'}, E_{i+k'}^{k-k'}, B_{i+k}^{j-k}) = 	\sum_{X_{i+k'}, C_i^{k'}, E_{i+k'}^{k-k'}, X_{i+k-1}, B_{i+k}^{j-k}} &\probm(X_{i+k'})
													\probm(C_i^{k'}, E_{i+k'}^{k-k'} | X_{i+k'}) \\
													&\probm(X_{i+k-1} | E_{i+k'}^{k-k'}, X_{i+k'})
                                                    \probm(B_{i+k}^{j-k} | X_{i+k-1})
\end{align*}

\paragraph{Complexity:}
The algorithm we propose needs to store the following quantities:
\begin{itemize}
\item $\probm(S_i^j | X_i, X_{i+j-1})$ for all $i, j, S$, which takes $\mathcal{O}(n^2 |G|^3)$ space.
\item $\probm(S_i^j, A_i^k, B_{i+k}^{j-k}, \widehat{C_j^k}(S,A,B))$ for all $i,j,k,S,A,B$ which takes at most $\mathcal{O}(n^3 |G|^3)$ space.
\end{itemize}
Therefore, we have a space complexity cubic in both the size of the sequence and the size of the grammar.
The run time, due to the summation involved is in $\OO(n^2 |G|^4)$ per variable, which gives a total complexity of $\OO(n^4 |G|^5)$.

\section{Conclusion}

We have investigated the complexity of imposing several hard constraints on Markov sequences at the sampling phase: binary equalities and grammar membership. We give a sketch of proof for the \#P-completeness of the binary equality constraint and identify three polynomial sub-cases.
We also give a proof for the \#P-completeness of the grammar membership constraint, and we identify two polynomial sub-cases, the unambiguous case and the new weakly-ambiguous case. Note that the subtle differences between perfect sampling, almost perfect sampling, exact counting and approximate counting could be investigated in more details to refine our statements.

\bibliographystyle{plain}
\bibliography{biblio}

\end{document}